\documentclass[aps,pre,twocolumn,showpacs,superscriptaddress]{revtex4-1}

\usepackage[T1]{fontenc}
\usepackage[latin1]{inputenc}
\usepackage[pdftex,dvipsnames,usenames]{color}
\usepackage{amsmath}
\usepackage{graphicx}
\usepackage{amssymb}

\bibstyle{apsrev.bib}

\makeatletter

\newcommand{\mean}[1]{{\left< #1 \right>}}

\usepackage{subfigure}
\usepackage{amsfonts}

\makeatother
\begin{document}

\title{Elastic lattice polymers}
\author{M. Baiesi}
\affiliation{Department of Physics, University of Padova, via Marzolo 8, 35131 Padova, Italy}
\affiliation{Institute for Theoretical Physics, K.U.Leuven, Celestijnenlaan 200D, B-3001 Leuven, Belgium}
\author{G.T. Barkema}
\affiliation{Institute for Theoretical Physics, K.U.Leuven, Celestijnenlaan 200D, B-3001 Leuven, Belgium}
\affiliation{Institute for Theoretical Physics, Utrecht University, Leuvenlaan 4, 3584CE Utrecht, The Netherlands}
\affiliation{Instituut-Lorentz, Universiteit Leiden, Niels Bohrweg 2, 2333 CA Leiden, The Netherlands}
\author{E. Carlon}
\affiliation{Institute for Theoretical Physics, K.U.Leuven, Celestijnenlaan 200D, B-3001 Leuven, Belgium}

\date{\today}

\begin{abstract}
We study a model of "elastic" lattice polymer in which a fixed number
of monomers $m$ is hosted by a self-avoiding walk with fluctuating
length $l$.  We show that the stored length density $\rho_m \equiv
1 - \langle l \rangle/m$ scales asymptotically for large $m$ as
$\rho_m=\rho_{\infty}(1-\theta/m + \ldots)$, where $\theta$ is the polymer
entropic exponent, so that $\theta$ can be determined from the analysis
of $\rho_m$.  We perform simulations for elastic lattice polymer loops
with various sizes and knots, in which we measure $\rho_m$.  The resulting
estimates support the hypothesis that the exponent $\theta$ is determined
only by the number of prime knots and not by their type. However, if
knots are present, we observe strong corrections to scaling, which help
to understand how an entropic competition between knots is affected by
the finite length of the chain.
\end{abstract}

\pacs{02.10.Kn, 36.20.Ey, 36.20.-r, 87.15.A-}

\maketitle

\section{Introduction}

According to renormalization group theory, the scaling properties
of critical systems are insensitive to microscopic details and are
governed by a small set of universal exponents~\cite{pelissetto02:RG}.
Also polymers can be considered as critical systems in the
limit where their length $l$ (the number of chained monomers)
diverges~\cite{dege79,desc90,vand98}.  For instance, the radius of gyration
of an isolated polymer in a swollen phase scales as $R_g \sim l^\nu$,
where $\nu \approx 0.587597(7)$~\cite{Clisby:2010:PRL} in $d=3$
dimensions is a universal critical exponent. One of the simplest models
in the universality class of swollen polymers is that of self-avoiding walks
(SAWs) on a lattice.
Hence, these have been used
extensively to extract information on critical exponents and scaling
functions~\cite{Clisby:2010:PRL,dege79,vand98,BFACF,nienhuis82,seno88:_theta,ishi89,madras90:_pivot,brak93,
grassberger95:theta2d,orla96,orla98,caracciolo98:_gamma,
jensen99,jens04,rechnitzer02:atmo,jans08,Clisby:2007:JPA,wolt06,heuk03,baiesi01:peculiar}.
The total number of SAWs, i.e.~their partition function, has the following
large-$l$ expansion
\begin{equation}
Z_l \sim \mu^l l^{\theta} \left(1 + A l^{-\Delta} + \ldots \right).
\label{Zl_saw}
\end{equation}
Here non-universal (model-dependent) quantities are the connectivity
constant $\mu$ and the amplitude of the corrections to scaling $A$.
The {\em entropic} exponent $\theta$ depends only on boundary conditions:
in  $d=3$ we have $\theta \equiv \gamma -1 = 0.1573(2)$~\cite{hsu04}
for an open chain whereas 
$\theta \equiv \alpha-2 = - d \nu = -1.762791(21)$~\cite{Clisby:2010:PRL}
for self-avoiding polygons (SAPs), that is, linear chains with the
two ends on adjacent lattice sites.  Renormalization group analysis suggests
that the exponent $\Delta$, characterizing the leading corrections to
the scaling behavior, is also universal \cite{pelissetto02:RG,desc90}

\begin{figure}[b] 
\includegraphics[width=4.5cm]{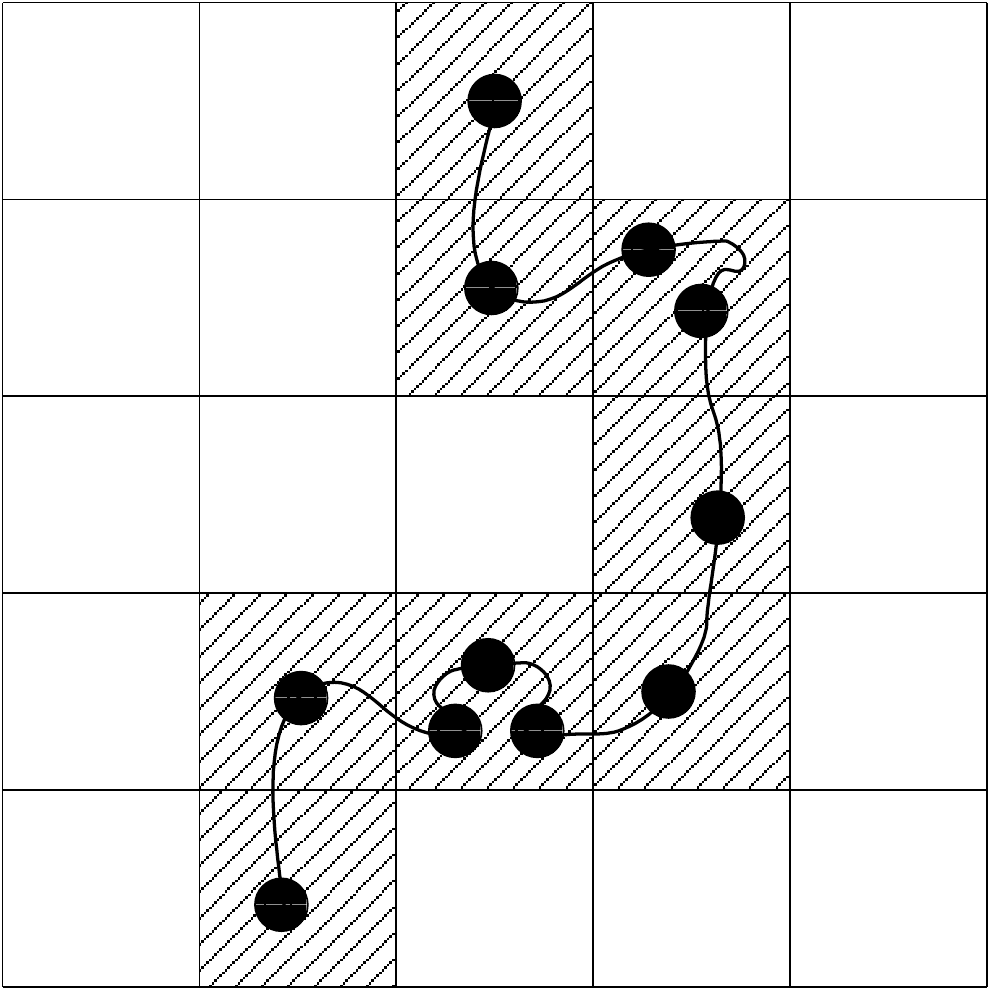}
\caption{Example of an elastic lattice polymer on a square lattice. 
This polymer is composed by $m=11$ monomers describing a SAW backbone of length
$l=8$ (dashed area).}
\label{fig01} 
\end{figure}

Models with full self-avoidance, such as SAPs, have been used to study the 
statistical properties of knotted chains~\cite{janse90,BOSTW:1997,orla96,orla98,
deguchi97,matsuda03:_aver,metzler02:_equil_shapes_flat_knots,zandi03,marcone05:_what,
orlandini07:rev,jans08,baiesi07:_ranking,Baiesi_et_al:2009:JCP,baie2010}.
Knots in polymers have attracted a lot of attention during the past 
years, also because of their occurrence in biopolymers as DNAs, RNAs and 
proteins~\cite{stasiak96:_elect_dna,rybenkov93:_probab,
Arsuaga:2005:Proc-Natl-Acad-Sci-U-S-A:15958528,taylor00:prot,lua06:knot_prot,
Burnier:2008:Nucleic-Acids-Res:18658246}.
As usual, SAWs represent a minimal effective model to grasp the essential, coarse-grained features
of polymer chains. Simulations of knotted SAPs in ensembles with fixed topology are
performed with a grand-canonical algorithm (BFACF~\cite{BFACF}, from the name of the authors) 
tuned to span a range of chain lengths 
(algorithms with fixed $N$ are not ergodic in this case). For this algorithm,
the tuning of step fugacities to $u\approx 1/\mu$ is necessary to achieve samplings of long chains.
It would be desirable to have a simpler and more stable method to sample the same chain lengths.

In this paper we study a class of polymers referred to as the 
{\em elastic lattice polymers} (ELPs), which are SAWs 
accumulating some stored length along their contour.
This leads in fact to a partial lifting of the self-avoidance
condition between consecutive monomers of an ELP, as sketched in
Fig.~\ref{fig01}. We will consider equilibrium properties of polymers with
a fixed number of monomers $m$, and in which as a consequence the length
$l\le m$ of the self-avoiding backbone described by the monomers
fluctuates. This explains the name ``elastic'', 
and implies a resemblance with the class of grand-canonical SAW models.

There are several reasons for studying this model. On the theoretical
side, it can be considered as an enhanced SAW: besides sharing critical
exponents with SAWs, its fluctuating length enables new avenues to
estimate critical exponents.
ELPs have been used in studies of polymer dynamics as phase separation
in polymer melts \cite{heuk03}, or in translocation through nanopores
\cite{wolt06}, but their equilibrium properties have so far received
little attention.

The key quantity we focus
on is the equilibrium averaged stored length density defined as
\begin{equation}
\rho_m \equiv 
\frac{m-\mean{l}}{m}.
\label{def_rho}
\end{equation}
where $\mean{l}$ depends on $m$.
As will be shown, $\rho_m$ has a simple asymptotic behavior for large
$m$ from which one can extract universal exponents: the leading correction
to the asymptotic value for $\rho_m$ scales as $\theta/m$, where $\theta$
is the entropic exponent defined by Eq.~(\ref{Zl_saw}). 
We illustrate the result of this approach for the case of ELPs with fixed knots.  
If knots are present, the stored length approaches its asymptotic value with strong,
knot-dependent corrections to scaling.  The expectation of a homogeneous
stored length within an equilibrated chain, combined with the knowledge
on how its density varies with the chain length, leads us to a new view
on the issue of entropic competition of knotted regions~\cite{zandi03}.
{
On the numerical side, we find that ELPs, compared to grand-canonical algorithms,
have the nice feature of stabilizing the sampling quite narrowly around an easily
tunable length $\mean{l}$.
}

This paper is organized as follows. In Section \ref{sec:theory}
we derive the expansion for $\rho_m$ as a function of $m$. 
In Section~\ref{sec:exact}, we illustrate how to estimate entropic exponents
via fits of $\rho_m$, with a reweighting of exact enumeration data 
for polymers on square and cubic lattices. 
In Section~\ref{sec:mc} we present Monte
Carlo simulations of ELPs containing a fixed knot
and determine the averaged stored length in equilibrium as a function
of $m$.  The entropic exponent $\theta$ is determined for different simple and double
knots. Finally, in Section~\ref{sec:compet2},
we discuss, on the basis of the obtained scaling behavior for $\rho_m$,
different possible scenarios for knot competitions.

\section{Scaling properties of the stored length}
\label{sec:theory}

Consider a polymer composed by $m$ monomers with lattice coordinates
defined by $\vec{r}_i$, $i=1,2,\ldots m$. Multiple occupancy of
neighboring monomers on the same lattice site means that we allow
configurations for which $\vec{r}_k = \vec{r}_{k+1} = \ldots =
\vec{r}_{k+p} = \vec{s}$. However if $\vec{r}_{k-1} \neq \vec{s}$ and
$\vec{r}_{k+p+1} \neq \vec{s}$, then no monomers other than those
of the interval $[k,k+p]$ are allowed to visit the site $\vec{s}$. The
lattice polymer so defined describes a self-avoiding backbone of length
$0 \leq l \leq m$. The two extremal cases are all monomers occupying the
same lattice point ($l=0$) and a fully stretched configuration without
multiple occupancy ($l=m$).
The equilibrium partition function for an ELP 
with $m$ monomers is given by
\begin{equation}
\widetilde{Z}_m = \sum_{l=0}^m \left( m \atop l \right) u^l Z_l ,
\label{zm}
\end{equation}
where the sum is over the length $l$ of the self-avoiding backbone and
$Z_l$ is the canonical partition function, which counts the number of
allowed configurations for the self-avoiding backbone, and whose asymptotic
is given in Eq.~(\ref{Zl_saw}). The factor $\left( m \atop l \right)$
in Eq.~(\ref{zm}) counts the number of ways the stored length can be
distributed over the backbone. For convenience an extra fugacity $u$
per site has been added.

Substituting Eq.~(\ref{Zl_saw}) in (\ref{zm}) and defining $\omega =
\mu u$, the average backbone length $\langle l \rangle$
can be computed from
\begin{equation}
\langle l \rangle = \omega \frac{\partial}{\partial \omega} 
\log \widetilde{Z}_m .
\label{getl}
\end{equation}

It is instructive to consider first the case of a partition function
of the type $Z_l = \mu^l$ in Eq.~(\ref{zm}), i.e. neglecting power-law
and correction to scaling terms in Eq.~(\ref{Zl_saw}). In this case
Eq.~(\ref{zm}) becomes
\begin{equation}
\widetilde{Z}_m ^{\infty} 
= \sum_{l=0}^m \left( m \atop l \right) \omega^l
= (1+\omega)^m .
\label{zm_theta0}
\end{equation}
Equation~(\ref{zm_theta0}) has the following interpretation: the
partition function for a walk of $m$ steps factorizes as each monomer
can either sit on the backbone (accumulating stored length with
weight $1$) or occupy a free site (with average weight $\omega$). From
Eq.~(\ref{getl}) we get the following value of the averaged backbone
length
\begin{equation}
l_\infty = \frac{m \omega}{1+\omega} .
\label{l0}
\end{equation}

We now go back to the full partition function in Eq.~(\ref{zm}).
For large $m$ and fixed $\omega$ the binomial factor is sharply
peaked around $l=l_\infty$. We approximate the binomial by a Gaussian
distribution as follows:
\begin{equation}
\left( m \atop l \right) \omega^l \approx (1+\omega)^m \frac 1 {\sqrt{2
\pi \sigma^2}} e^{-(l-l_\infty)^2/2\sigma^2},
\end{equation}
where
\begin{equation}
\sigma^2 = \frac{m \omega}{(1+\omega)^2}.
\label{sigma}
\end{equation}
The Gaussian approximation differs from the binomial by terms which are
exponentially small for large $m$, which are of higher order in the
large-$m$ expansion we are interested in, so they can be safely neglected.
We replace now the discrete sum in Eq.~(\ref{zm}) by an integral over all 
lengths, extending the domain of integration in the whole real axis:
\begin{equation}
\widetilde{Z}_m = \frac{(1+\omega)^m}{\sqrt{2 \pi \sigma^2}}
\int_{-\infty}^{+\infty} dl \ e^{-(l-l_\infty)^2/2\sigma^2} l^\theta
\left(1 + A l^{-\Delta} \right)
\label{zm_saddle}
\end{equation}
where we have replaced the asymptotic form of $Z_l$ as given in
Eq.~(\ref{Zl_saw}).
The replacement of the sum by an integral brings corrections in
Eq.~(\ref{zm_saddle}) which are of higher order in $1/m$ and for
our purposes can be neglected.

We solve the integral in Eq.~(\ref{zm_saddle}) by using a saddle point
approximation. A simple rescaling $l=x \ l_\infty$ gives
\begin{equation}
\widetilde{Z}_m = (1+\omega)^m \sqrt{\frac{m \omega}{2 \pi}} 
\int_{-\infty}^{+\infty} dx \ e^{m \omega \Gamma(x)}
\label{gauss_int2}
\end{equation}
with 
\begin{equation}
\Gamma (x) = \frac{-(x-1)^2}{2} + 
\frac{\theta \log (x\ l_\infty) + 
\log \left(1 + A (x\ l_\infty)^{-\Delta} \right)}
{m \omega} 
\label{gammam}
\end{equation}
Let $\bar{x}$ the maximum of $\Gamma (x)$. We have:
\begin{eqnarray}
\widetilde{Z}_m &\approx&  (1+\omega)^m \sqrt{\frac{m \omega}{2 \pi}}
e^{m \omega \Gamma(\bar{x})} 
\sqrt{\frac{2 \pi}{m \omega \left| \Gamma''(\bar{x})\right|}}
\nonumber \\ 
&=& (1+\omega)^m 
\frac{e^{m \omega \Gamma(\bar{x})}}{\sqrt{\left| \Gamma''(\bar{x})\right|}}.
\label{final_saddle}
\end{eqnarray}

Equation~(\ref{gammam}) implies that the maximum of $\Gamma(x)$ 
in the large-$m$ limit is $\bar{x} = 1 + {\cal O}(1/m)$, giving 
\begin{equation}
\left|\Gamma'' (\bar{x})\right| = 1 + {\cal O}\left(\frac{1}{m}\right),
\label{gamma2}
\end{equation}
which produces higher-order terms, which we neglect in the large-$m$ expansion.
In addition:
\begin{equation}
m \omega \Gamma(\bar{x}) = 
\theta \log l_\infty + 
\log \left(1 + A l_\infty^{-\Delta} \right) + \ldots
\label{m_omega_gamma}
\end{equation}
Equations~ (\ref{final_saddle}), (\ref{gamma2}) and (\ref{m_omega_gamma})
again show that the leading contribution to the partition function
$\widetilde{Z}_m$ is $(1+\omega)^m$, 
{
but also that the subleading contribution $\sim l_\infty^\theta\sim m^\theta$ 
has the same entropic exponent $\theta$ of SAWs.
} 
From Eq.~(\ref{getl}) we get
\begin{equation}
\langle l \rangle = l_\infty  \left[ 
1 + \frac{\theta}{ m \omega}
- \frac{A}{m \omega} \left( \frac{1+\omega}{ m \omega}\right)^\Delta 
\right]
\end{equation}
and the stored length density (\ref{def_rho})  becomes
\begin{equation}
\rho_m = \rho_{\infty} \left[ 1 - \frac{\theta}{m} 
+ \left( \frac{1+\omega}{ \omega}\right)^\Delta  
\frac A {m^{1+\Delta}}\right] ,
\label{rho}
\end{equation}
where we defined
\begin{equation}
\rho_{\infty} = \frac{1}{1+\omega} .
\label{rho_infty}
\end{equation}

The expansion (\ref{rho}) is valid provided $\Delta < 1$. The neglected terms
coming from the replacement of the sum with an integral, and from the
Gaussian integration in Eq.~(\ref{gauss_int2}), are of the order $1/m^2$
(except if $\Delta > 1$, the ${\cal O}(1/m^2)$ terms would dominate over the
$1/m^{1+\Delta}$.)  The value of the exponent $\theta$ can then be obtained from
a plot of $\rho_m$ vs.~$1/m$, as the slope in the limit $1/m \rightarrow 0$.
Using the high-precision literature values for the connectivity constants $\mu$,
one obtains a very accurate estimate of $\rho_\infty$.

With the definition of $\rho_\infty$ in (\ref{rho_infty}) we can rewrite
the variance (\ref{sigma}) as
\begin{equation}
\sigma^2 = m\, \rho_\infty(1-\rho_\infty)
\end{equation}
This form reveals clearly that the largest $\sigma$ for a given $m$ 
is achieved with $\rho_\infty=1/2$,
i.e.\ with a fugacity $u = \mu^{-1}$. We can think of this regime as the 
maximally elastic one.
{
In all cases, note that the relative polydispersity $\sigma / m$ 
of the chains goes to zero $\sim m^{-1/2}$ for $m\to \infty$, hence
the chain lengths $l$ are narrowly distributed around their average $\mean{l}$.
This allows us to use saddle-point approximations (see Sect.~\ref{sec:compet2}) 
and leads to metric properties in the universality class of SAWs 
(e.g. radius of gyration scaling as $\sim m^\nu\sim\mean{l}^\nu$).
Hence, ELPs share both exponents $\theta$ and $\nu$ with SAWs.
}
\section{Stored length from exact enumerations data}
\label{sec:exact}

As a first illustration of the scaling behavior of the stored length
$\rho_m$ as a function of the number of monomers $m$, we consider exact
enumeration data for SAWs and SAPs on square and cubic
lattices, which are taken from the published literature \cite{jens04}.
Enumeration techniques provide exact values for the total number
of SAWs $Z_l$ as a function of their length
$l$. We use these values for $Z_l$ to compute $\widetilde{Z}_m$
from Eq.~(\ref{zm}). The stored length $\rho_m$ is obtained from the
average $\langle l \rangle$, using Eq.~(\ref{def_rho}). We have the freedom 
to choose the value of the fugacity $u$ in Eq.~(\ref{zm}).

\begin{figure}[t]
\includegraphics[width=8cm]{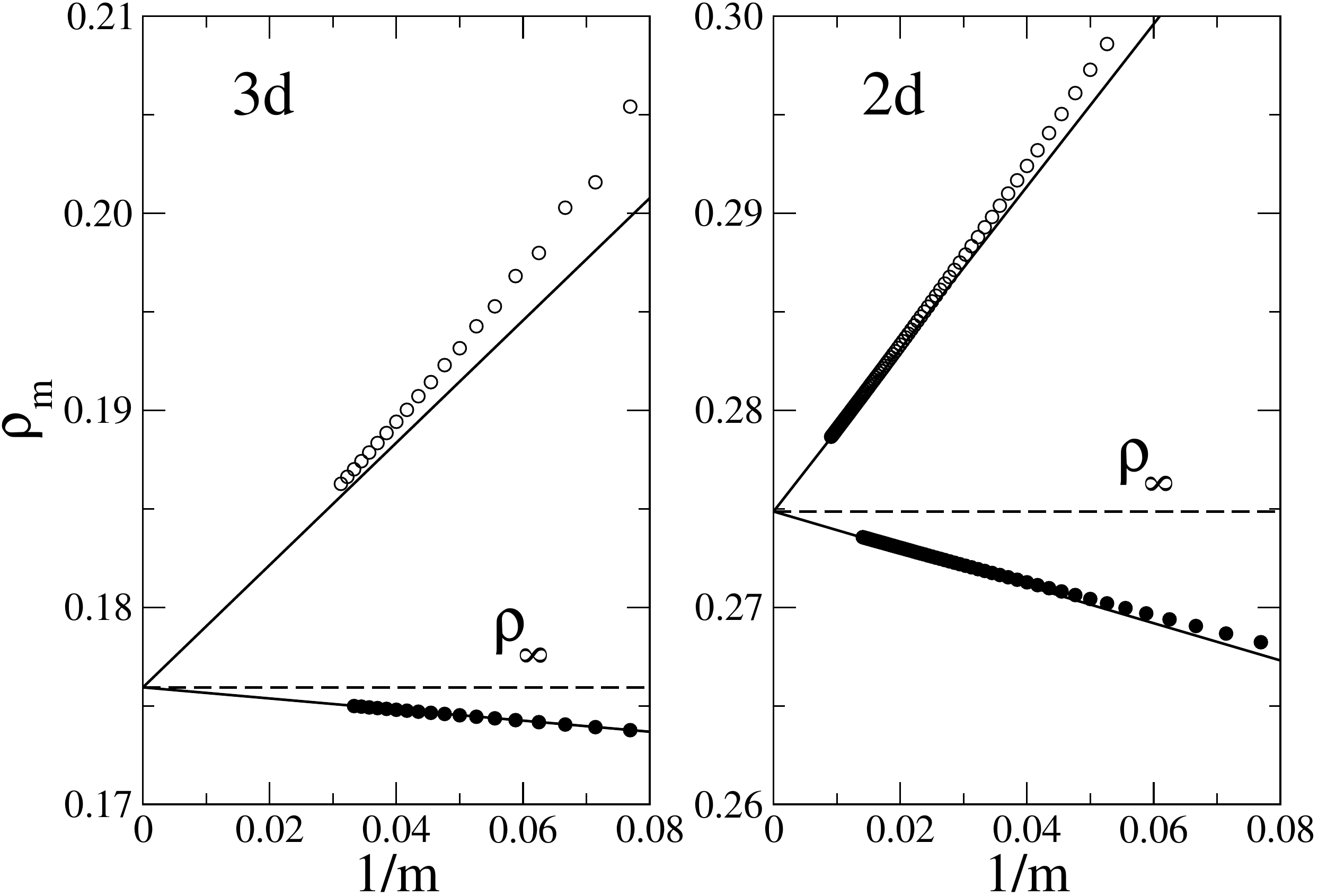}
\caption{Stored length calculated from exact enumeration data for
SAWs (bullets) and SAPs (empty circles) on the cubic lattice ($d=3$) 
and on the square lattice ($d=2$). The
solid lines are the leading terms in the $\rho_m$ vs.~$1/m$ expansion when
using the expected value for the exponent $\theta$.  The asymptotic values
$\rho_\infty = 1/(1+\mu)$ for cubic and square lattices are shown as a
dashed line.
}
\label{fig02}
\end{figure}

Figure~\ref{fig02} shows a plot of $\rho_m$ as a function of $1/m$ for
three and two dimensions, obtained by setting $u=1$ in Eq.~(\ref{zm}). The
data converge to the expected asymptotic value, which is 
$\rho_\infty \simeq 0.175931$ (cubic)  and 
$\rho_\infty \simeq 0.2748643$ (square). 
These are obtained from Eq.~(\ref{rho_infty}) with  $\omega = u \mu = \mu$ and
the following values for the connectivity constants: 
$\mu = 4.684044(11)$~\cite{Clisby:2007:JPA} (cubic) and 
$\mu = 2.63815852927(1)$ (square)~\cite{jensen99}.      

The solid lines in Fig.~\ref{fig02} are the linear terms in the expansion
of Eq.~(\ref{rho}) where the value of $\theta$ is that for open walks
($\theta = 11/32$ in $d=2$ and $\theta \simeq 0.157$ in $d=3$) and polygons
($\theta = -3/2$ in $d=2$ and $\theta \simeq -1.76$ in $d=3$).  The results show
that the linear scaling in $1/m$ sets in already for short polymers
($m \approx 20$). In addition we observe that the corrections to the
leading scaling behavior are stronger for closed walks (empty circles)
compared to the open walks case (filled circles).

\begin{table}[!b]
\caption{Summary of the exponents obtained from the extrapolation of the 
approximants $\theta_m$ defined in Eq.~(\ref{thetam}). The data are for
SAWs  and SAPs~\protect{\cite{jens04}}.
The last column gives the exact two dimensional data~\protect{\cite{vand98}}.}
\vskip 0.2cm
\begin{tabular}{|c|c|c|c|}
\hline
type    & max $m$ \protect{\cite{jens04}} & $\theta$ & $\theta_{\rm ex}$ \\
\hline
SAW, $d=2$& 71	& $0.3437(2)$ & $0.34375 (=\frac{11}{32})$\\
SAP, $d=2$& 110	& $-1.500(1)$ & $-\frac 3 2$\\
\hline
SAW, $d=3$& 30	& $0.158(2) $ & -- \\ 
SAP, $d=3$& 32	& $-1.75(2)$  & -- \\
\hline
\end{tabular}
\label{table_exact}
\end{table}

We performed finite-size extrapolations to obtain estimates of $\theta$
from $\rho_m$. The two-dimensional data have been extrapolated by means of
the Burlisch-Stoer (BST) algorithm \cite{henk88}, using the finite-$m$
approximants
\begin{equation}
\theta_m \equiv \frac{\rho_m - \rho_{m-1}}{\rho_{m-1}/m - \rho_{m}/(m-1)}
\label{thetam}
\end{equation}
which are the ratios between slope and intercept of the line joining the
points $(1/(m-1),\rho_{m-1})$ and $(1/m,\rho_{m})$, 
i.e.~they are finite-size estimates of the ratio between
$\rho_\infty \theta$ and $\rho_\infty$.
The BST algorithm starts
with a sequence of $N$ elements, and generates iteratively sequences
of $N-1$, $N-2$ $\ldots$ elements which are expected to converge faster
at each iteration step.  It involves a free parameter
($\Omega$), which roughly measures the effective leading correction
exponent.
In our extrapolations, an optimal value of $\Omega$ was selected requiring
a minimal standard deviation of the last five sequences generated by the
iterative algorithm.  The extrapolations were repeated for different
values of the fugacity parameter $u$ and the error was estimated from the
variation on these values.  For three-dimensional data, the BST algorithm
turned out not to be very accurate, particularly for loops. The reason is
that $\rho_{m}$ for small $m$ has some subleading oscillatoric behavior
which is not sufficiently damped during the BST iterations. The result
is that the accuracy of the extrapolation is poor.  For these data we
use instead a non-linear fit, fixing $\rho_\infty$ and keeping $\theta$,
$A$ and $\Delta$ as fitting parameters.

The extrapolated values for $\theta$ are reported in
Table~\ref{table_exact}; these are accurate and in good agreement with
exact data in two dimensions and also with the best numerical estimates in
three dimensions ($\theta = 0.1573(2)$ \cite{hsu04} for walks, 
$\theta = - 1.76279(2)$ for polygons --- assuming 
hyperscaling $\alpha-2 = -d \nu$ 
with $\nu=0.587597(7)$~\cite{Clisby:2010:PRL}), 
which shows that reliable values of the entropic
exponents can be extracted from the scaling of the stored length.

\begin{figure}[t]
\includegraphics[width=3.2cm]{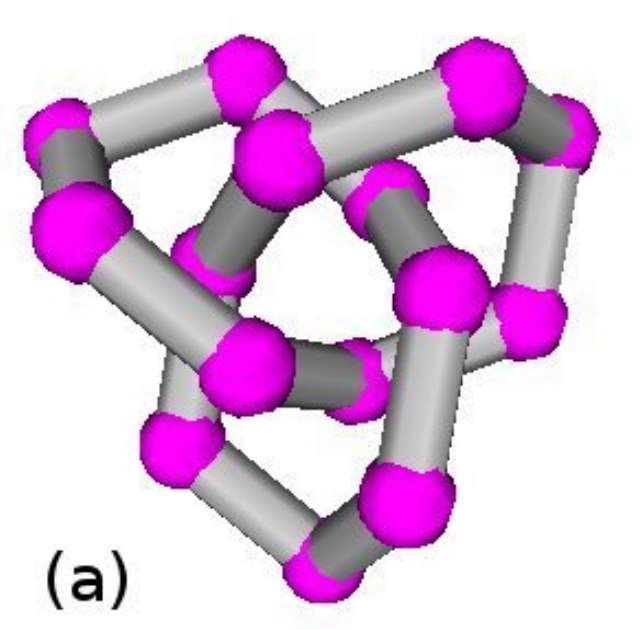}
\includegraphics[width=4.8cm]{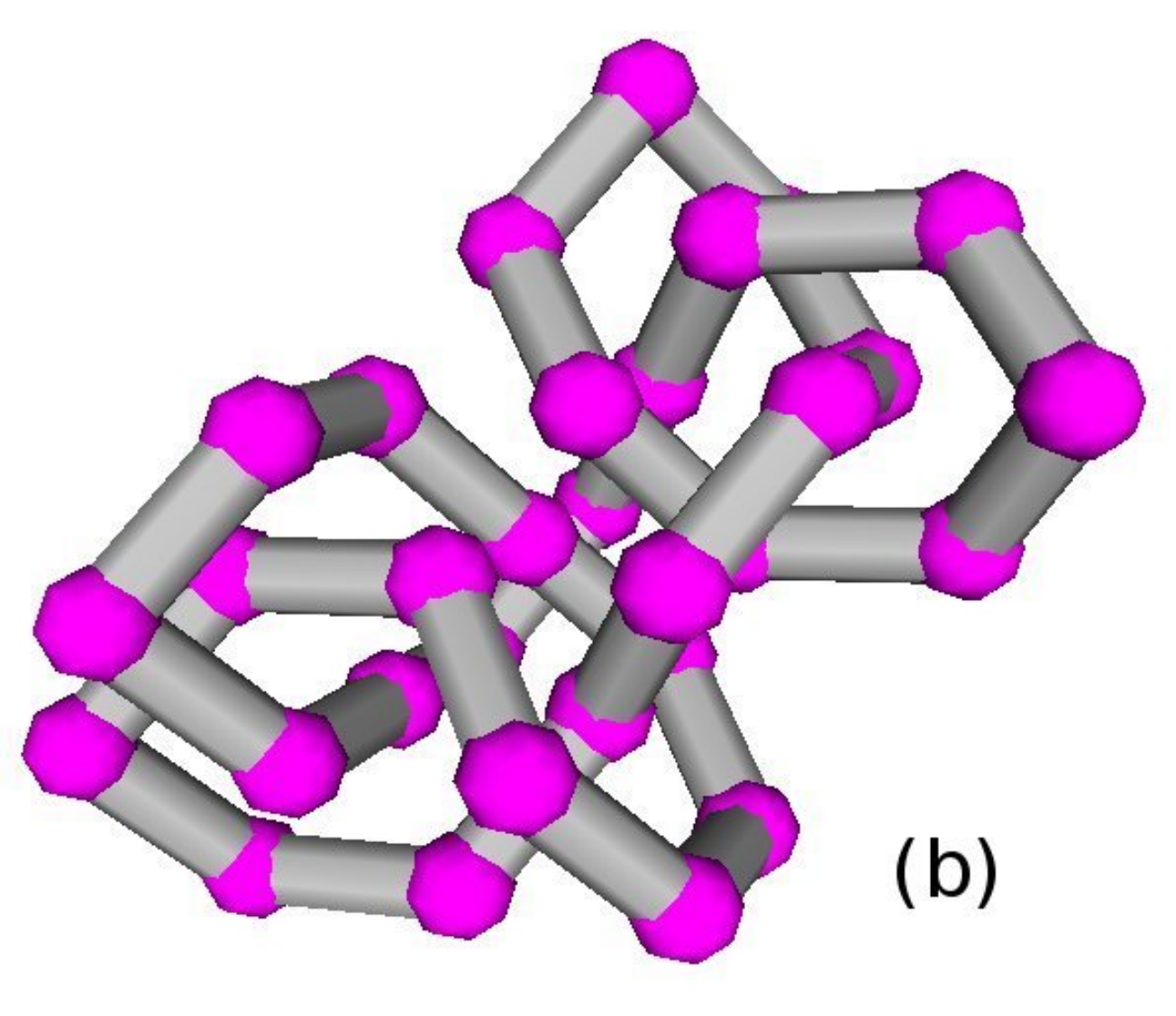}
\caption{(Color online) Examples of knots on the fcc lattice: (a) a $3_1$ knot (trefoil)
with $l=15$ steps and (b) a composite $3_1\#4_1$ knot with $l=31$ steps
(the notation $k_1\#k_2$ indicates a closed polymer ring 
with two knots, one of type $k_1$ and one of type $k_2$).
These configurations have been used as backbones with stored length $m-l$
for starting the simulations of ELP with  $m\ge l$ monomers.}
\label{fig:ini}
\end{figure}

\section{Entropic exponents of knotted polymers}
\label{sec:mc}

We now turn to the study of equilibrium properties of ELP rings 
with some fixed topology.
Here we will show how the knowledge of the stored length $\rho_m$ 
can be exploited to investigate equilibrium properties of knotted polymers.

\begin{figure}[t]
\includegraphics[width=8cm]{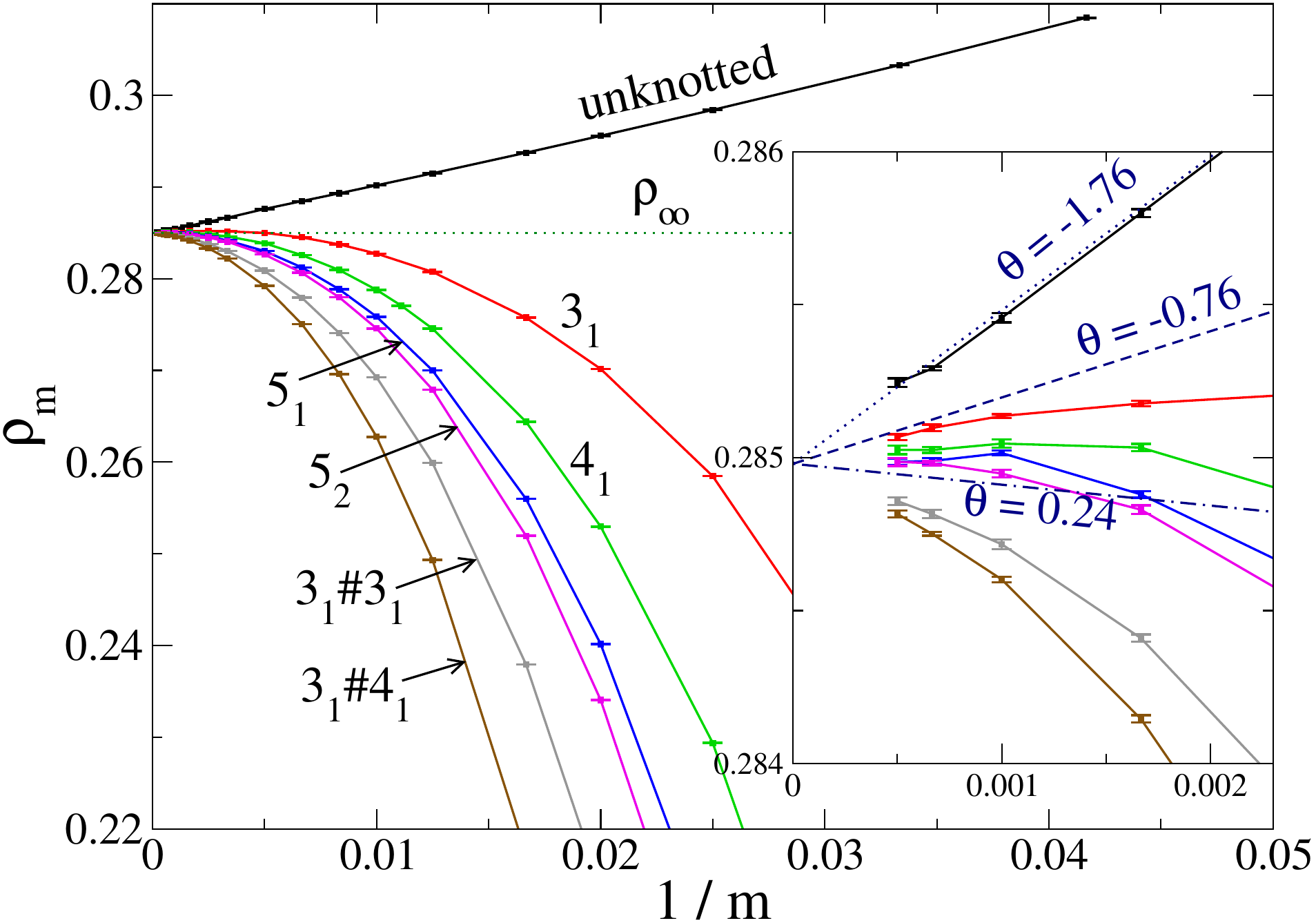}
\caption{(Color online) Plot of the average equilibrium stored length density $\rho_m$
as a function of the inverse monomer number $1/m$ for closed polymers with
some fixed topology. The simulations were extended to polymers of lengths
up to $m=2000$. From top to bottom the data refers to: unknotted ring,
$3_1$, $4_1$, $5_1$ and $5_2$ knots. The two bottom data set correspond
to configuration with two knots: $3_1\#3_1$ and $3_1\#4_1$, respectively.
The horizontal dashed line is $\rho_\infty=0.28498$, as expected from
Eq.~(\ref{rho_infty_model}). 
Inset: zoom of the asymptotic region. Straight lines represent
$\rho_{\infty}(1-\theta/m)$:
the dotted line corresponds to the conjectured value $\theta = - 3 \nu$,
the dashed line to $\theta = - 3 \nu+1$, and the dot-dashed line to
$\theta = - 3 \nu+2$.
}
\label{fig04}
\end{figure}

We have performed Monte Carlo simulations of ELPs on the
face-centered-cubic (fcc) lattice, with an algorithm that was recently
used to study translocation dynamics~\cite{wolt06} and phase separation
in polymer melts~\cite{heuk03}.  The allowed Monte Carlo update moves
include reptation, i.e.~the diffusion of stored length along its backbone
and Rouse-like moves which locally change the backbone configuration.
(For more details see Ref.~\cite{heuk03}).

The setup of the simulation is as follows. We start from a backbone
with a minimal number of steps on the fcc lattice, as those shown in
Fig.~\ref{fig:ini}. A total number of monomers $m$ are distributed
randomly over this backbone. These configurations are then relaxed
to equilibrium. Typically $m$ is much larger than the initial length
(we simulated polymers with $m$ up to $2000$) so that relaxation to
equilibrium corresponds to an expansion of the backbone. The Monte
Carlo moves preserve the knot topology imposed initially. Once
equilibrium is reached we start the sampling of the stored length
density $\rho_m$.

An additional weight (equal to $4$) is introduced for moves that
accumulate monomers on the same lattice point, which corresponds to a
fugacity factor $u=1/4$ in Eq.~(\ref{zm}). This leads to the following
asymptotic value for the stored length density:
\begin{equation}
\rho_\infty = \frac 1 {1+ \mu/4} = 0.28498(1)
\label{rho_infty_model}
\end{equation}
where the numerical value is obtained by considering the most accurate
available estimate $\mu = 10.0362(6)$ \cite{ishi89} for the connectivity
constant of SAWs on fcc lattice~\footnote{In principle 
we should consider the connectivity constant $\mu_0$ 
of the ensemble of configurations with a given knot, 
which is slightly smaller than $\mu$.  However, their estimated 
difference~\cite{orla2010}  is smaller than the numerical error for $\mu$.}.

Figure~\ref{fig04} shows the scaling behavior of $\rho_m$ as a function
of $1/m$ for an unknotted polymer ring, for single and double knots.  All
data converge asymptotically to the value $\rho_\infty$ obtained from
Eq.~(\ref{rho_infty_model}). This value is shown as a dashed horizontal
line in Fig.~\ref{fig04}.  We note that the approach to $\rho_\infty$ of
the numerical data for unknotted rings is quite different for those of
rings with knots (a detail of the asymptotic region is shown in the
inset of Fig.~\ref{fig04}): the data for the unknotted topology approach
the asymptotic value with a clear $1/m$ scaling behavior.
For topologies with knots instead there is a pronounced curvature in 
the $\rho_m$ vs.~$1/m$ plot, deriving from strong corrections to scaling.
These corrections are stronger for an increasing knot complexity
and for an increasing number of knots.
{
The shortest length $l_{\min}$ of a knot on a lattice is a good indicator of its complexity,
and in this model for $m=l=l_{\min}$ by definition the chain can only be fully stretched, 
i.e.~$\rho_{m_{\min}}=0$.
Our data show that the crossover from this initial topological stretching to the asymptotic regime $\sim m^{-1}$
grows quickly with the value of $m_{\min}$.
In this view, the fact that for unknotted chains on the fcc lattice
one has $m_{\min}=3$, much smaller than that of the simplest knot (the trefoil with $m_{\min}=15$)
explains why corrections to scaling are negligible for unknotted chains.
}

\begin{table}[!b]
\caption{Summary of the estimated entropic exponents obtained from the
scaling behavior of the stored length with a three parameters fit ($\theta$,
$\Delta$ and $A$ in Eq.~(\ref{rho})). The asymptotic value $\rho_\infty$
is kept fixed.  The last column shows the range of polymer sizes used
in the fit.}
\vskip 0.2cm
\begin{tabular}{|c|c|c|c|}
\hline
Knot type & $\theta$ & $\Delta$ & range of $m$\\
\hline
unknotted & $-1.76(3)$	& --- 		& $\geq 200$\\
\hline
$3_1$ 	& $-0.75(5)$	& $0.5$-$0.7$ 	& $\geq 400$\\
$4_1$ 	& $-0.6(1)$   	& $0.9$-$1.2$ 	& $\geq 300$\\
$5_1$ 	& $-0.5(2)$    	& $0.9$-$1.2$ 	& $\geq 300$\\
$5_2$ 	& $-0.3(3)$    	& $0.9$-$1.2$ 	& $\geq 300$\\
\hline
$3_1\#3_1$ & $0.4(2)$ 	& $1.2$-$1.4$ 	& $\geq 300$\\
$3_1\#4_1$ & $0.8(2)$ 	& $1.2$-$1.4$ 	& $\geq 300$\\
\hline
\end{tabular}
\label{tableII}
\end{table}

We estimated the entropic exponent $\theta$ using the scaling behavior
predicted by Eq.~(\ref{rho}).  In the unknotted case due to the manifest
absence of curvature of the data, we restricted ourselves to a linear
fit setting $A=0$ and $\rho_\infty = 0.28498$ in Eq.~(\ref{rho}) and
using $\theta$ as the only free parameter. The fit, restricted to $m
\geq 200$, yields $\theta = -1.76(3)$, confirming that the entropic
exponent for rings with fixed unknotted topology is identical to that
for SAPs with no topological constraints.

A closer look at the data reveals that the stored length density for knots
$3_1$, $4_1$ and $5_1$ is non-monotonic.  As the data asymptotically
approach $\rho_\infty$ from above, Eq.~(\ref{rho}) implies a negative
value of the exponent $\theta$. We performed a non-linear three-parameters
fit to the data based on Eq.~(\ref{rho}): $\theta$, $\Delta$ and $A$
are fitting parameters while we fix $\rho_\infty=0.28498$, as predicted
by Eq.~(\ref{rho_infty_model}).
The results of the non-linear fits are given in Table~\ref{tableII}. The
estimated exponent changes sign from single knot ($\theta < 0$) to double
knots ($\theta > 0$). A range of correction-to-scaling exponents $\Delta$
providing optimal fits were selected and these are given in the third
column of Table~\ref{tableII}. Error estimates for $\theta$ reflect the
variability in $\theta$ from the different values of $\Delta$ used in
the analysis. For the knots studied, the most accurate estimate for $\theta$
is that of the $3_1$ knot, yielding $\theta = -0.75(5)$. The error increases with
the knot complexity.  For single knots we also note a change in the range of
correction-to-scaling exponents from $\Delta \approx 0.6$ for the $3_1$
knot to $\Delta \approx 1.1$ in the other knots.

\begin{figure}[!tb]
\begin{center}
\includegraphics[angle=0,width=8cm]{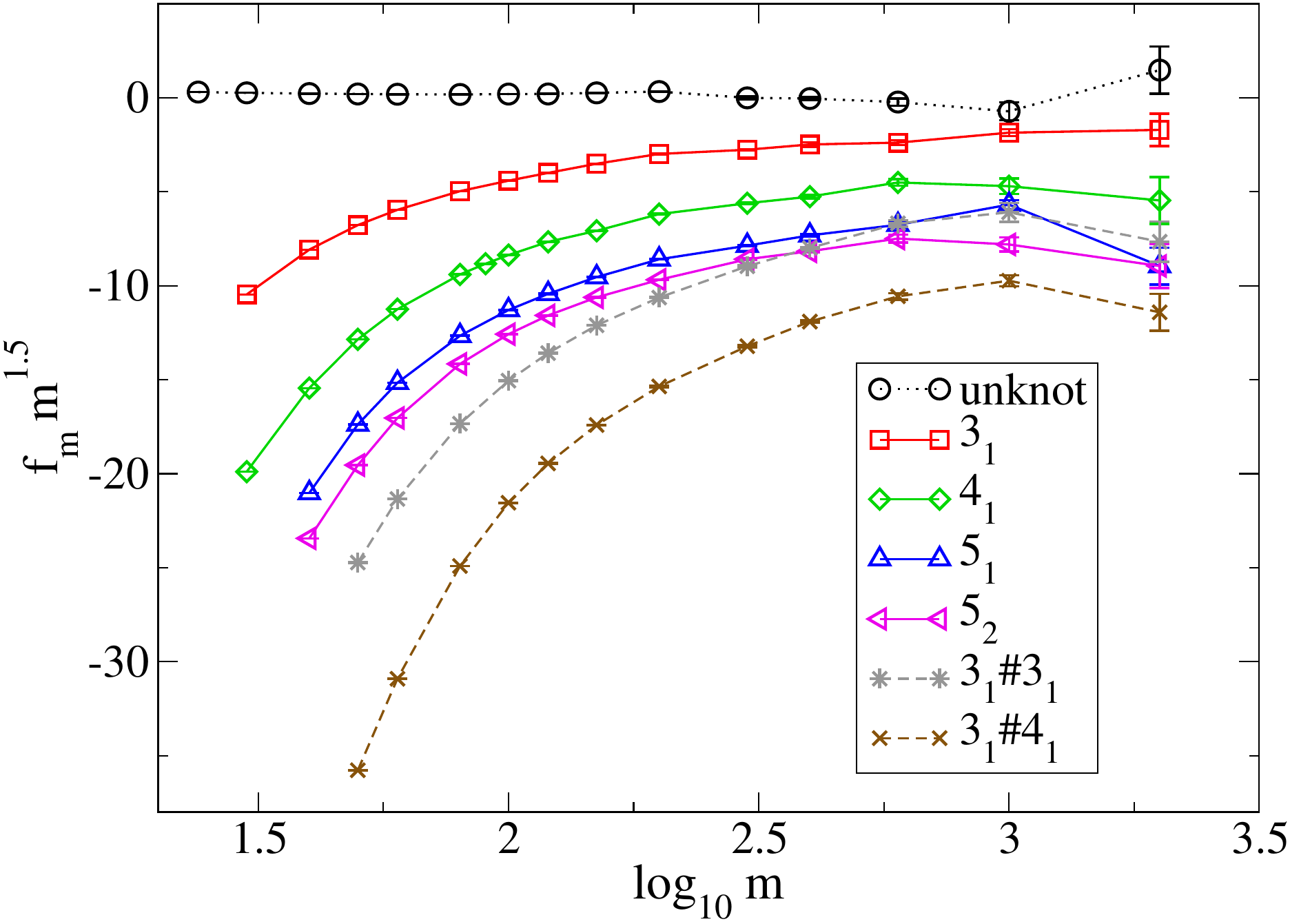}
\end{center}
\caption{(Color online) Plot of $f_m m^{1+\Delta}$  vs.~$\log_{10} m$, with $\Delta=1/2$.
The data tend to a constant for large $m$, which, as discussed in the text, is
consistent with a correction-to-scaling exponent of $\Delta \approx 0.5$.}
\label{fig:cs}
\end{figure}

It has been suggested \cite{orla96,orla98,baie2010} that for a knot
$k$ with $\pi_k$ prime components, the entropic exponent is given
by $\theta_k = \theta + \pi_k$, where $\theta$ is the exponent for a
polymer ring without fixed topology.  If this is the case we expect for
a single knot an exponent $\theta = -3 \nu +1 = -0.76$ while for double
knots $\theta = -3 \nu + 2 = 0.24$ (these conjectured values are shown
as dashed lines in the inset of Fig.~\ref{fig04}). The idea behind this
suggested scaling is that localized knots are like sliding entities, which can
occupy any of the $l$ sites of a chain, thus contributing entropically
with a factor $l$ in the partition function.  Our numerical
results fully support this conjecture for the single $3_1$ knot and
also for the double $3_1\#3_1$ knot. Results for the other knots seem
to overestimate $\theta$ with respect to the conjectured values. It is
likely that the deviations from the conjectured values are due to strong
finite-size effects. An indication of this is the value of the correction-to-scaling
exponent obtained from the fits, which, with the exception
of the $3_1$ knot, is estimated as $\Delta \approx 1$. Renormalization
group arguments~\cite{pelissetto02:RG} for magnetic $O(N)$ models, which
map into polymer models in the limit $N \to 0$ \cite{dege79}, predict
instead $\Delta \approx 0.55$, and Clisby~\cite{Clisby:2010:PRL} finds 
$\Delta \approx 0.528(12)$ in simulations of very long SAWs
(this is in agreement with the range of values obtained 
in the extrapolations of the numerical data for the $3_1$
knot, see Table~\ref{tableII}). We also remark that a value $\Delta > 1$
is at odds with the expansion of the stored length of Eq.~(\ref{rho})
in which it was implicitly assumed $\Delta < 1$, the $1/m^{1+\Delta}$
term would be otherwise dominated by ${\cal O}(1/m^2)$ corrections, which
were neglected in the computation of $\rho_m$ leading to Eq.~(\ref{rho}).

\begin{figure*}[t]
\begin{center}
\includegraphics[angle=0,width=6cm]{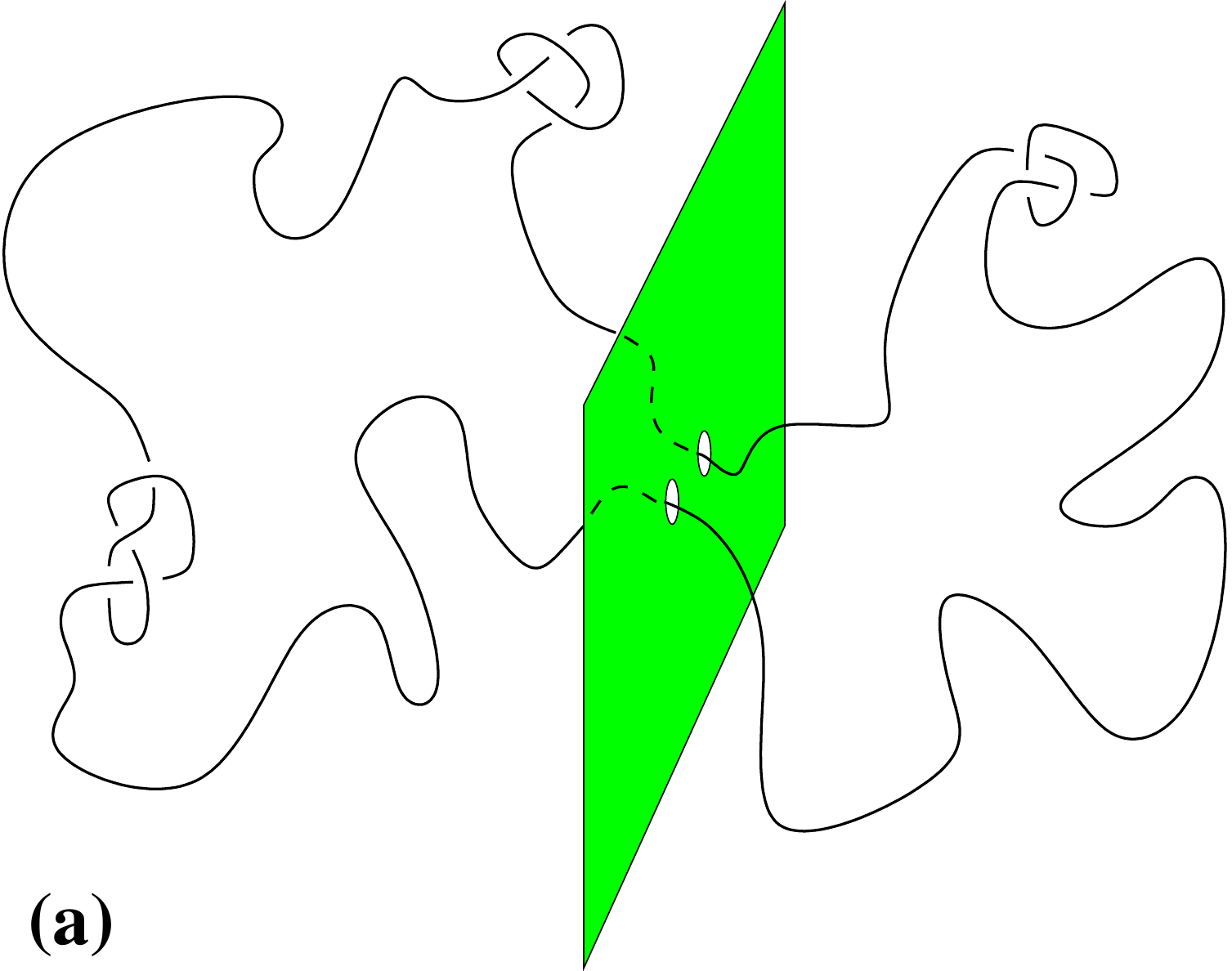}
\hskip 12mm
\includegraphics[angle=0,width=6.8cm]{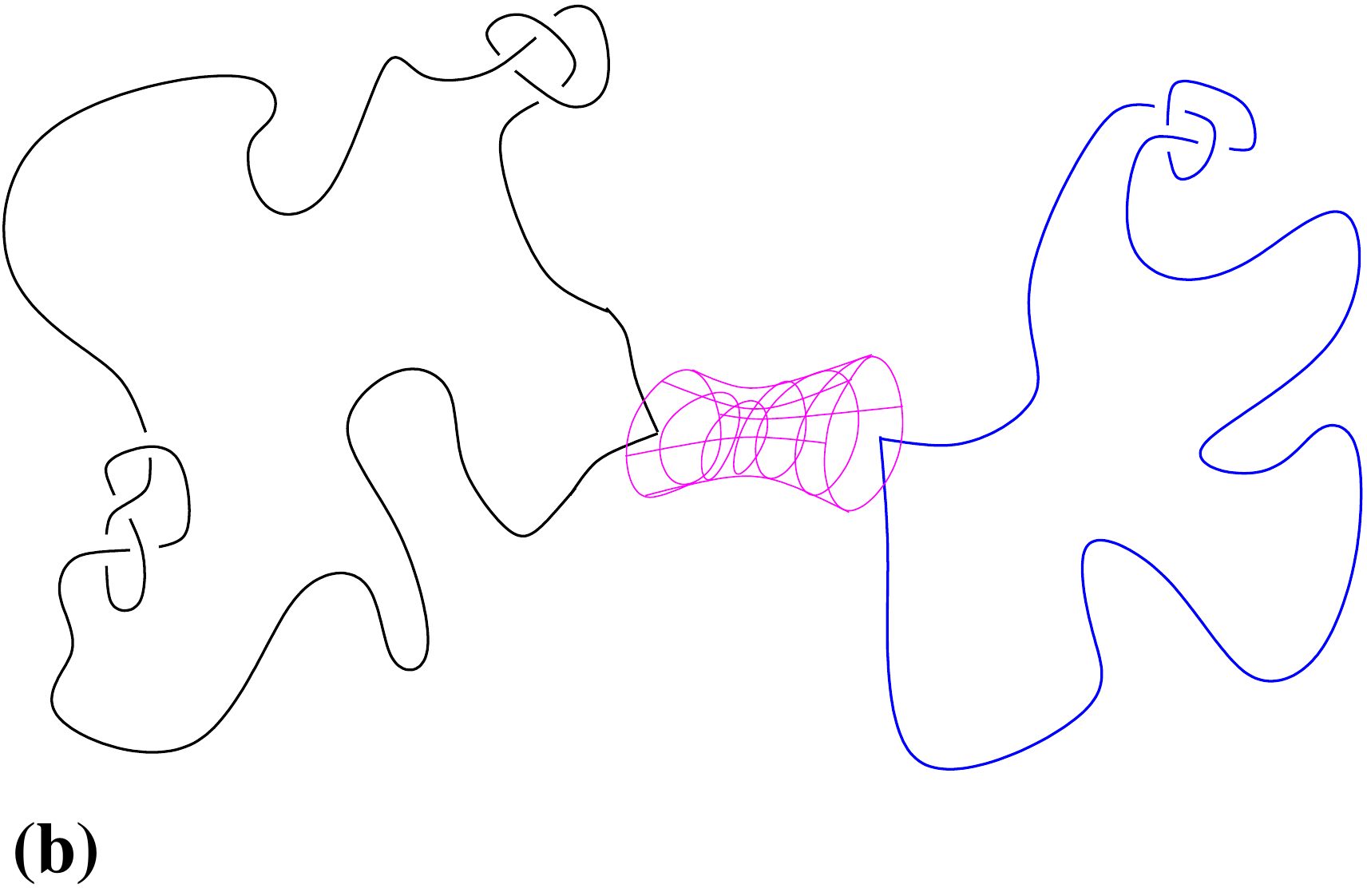}
\end{center}
\caption{(Color online) (a) Sketch of an entropic competition: a portion of the ring
polymer, with two knots, is constrained to stay on the left half-space
(holes are small enough to forbid more than one monomer at a time to
pass), the remaining part has one knot and is on the other side of the
wall. The total length $N$ of the chain is constant but the lengths
$m$ and $N-m$ of the two subchains can fluctuate.  (b) The virtual
version (without the wall) of the same competition: the two polymers
swim in separate dilute solutions and are coupled via a ``wormhole''
trough which they can exchange a monomer (hence not a knot) at a time.
\label{fig:sketch}}
\end{figure*}

In the sequel we fix the entropic exponents to the conjectured values 
$\bar \theta \equiv -3 \nu + \pi_k$ and subtract from $\rho_m$ the constant
and leading correction in $1/m$ as 
\begin{equation}
f_m \equiv \rho_m - \rho_\infty\left(1-\frac{\bar \theta}{m}\right).
\end{equation}
For this quantity we expect the following scaling behavior
\begin{equation}
f_m \simeq \frac{\overline A}{m^{1+\Delta}} + \frac{\overline B}{m^{2}},
\label{fms}
\end{equation}
where a next-order $1/m^2$ term has been added.

\begin{table}[!b]
\caption{Fits of $\overline A$ and  $\overline B$, assuming $\Delta=1/2$
in Eq.~(\ref{fms}).}
\vskip 0.2cm
\begin{tabular}{|c|c|c|}
\hline
Knot type & $\overline A$ & $\overline B$ \\
\hline
$3_1$ 	& $-0.78$	& $-33$ \\
$4_1$ 	& $-1.9$	& $-63$ \\
$5_1$ 	& $-3.5$        & $-73$ \\
$5_2$ 	& $-4.1$        & $-80$ \\
\hline
$3_1\#3_1$ & $-1.4$ 	& $-130$ \\
$3_1\#4_1$ & $-3.3$ 	& $-173$ \\
\hline
\end{tabular}
\label{tableIII}
\end{table}

Figure \ref{fig:cs} plots $f_m m^{1+\Delta}$, where we set $\Delta = 0.5$, 
as a function of $m$.  The fact that this quantity approaches
a constant value for large $m$ supports an estimate of the
correction-to-scaling exponent $\Delta \approx 0.5$, as expected for swollen
polymers~\cite{vand98}. In addition the constant $\overline A$ is negative
and its magnitude quickly increases with knot complexity. This is
also visible in Fig.~\ref{fig04} as the effect of increasing $A$ is
that of producing an increased curvature in a plot of $\rho_m$ vs.~$1/m$.
It is perhaps not surprising that finite-size effects increase with the
knot complexity, as more complex knots are expected to occupy a larger
portion of the polymer.
In table~\ref{tableIII} we list our estimates of  $\overline A$ and
$\overline B$, obtained by means of linear fits to data in the form
$f_mm^{1.5}$ vs.~$m^{-0.5}$.
The values of $\overline B$ are almost
two orders of magnitude larger than those of $\overline A$, explaining
the fitted (effective) leading exponent $\Delta\approx 1$.

\section{Knots competition}
\label{sec:compet2}

In this Section we discuss entropic competition between knotted
polymers in the context of ELPs. The idea of entropic competition
between polymers with various constraints was introduced 
in Ref.~\cite{zandi03} as a direct way to estimate polymer entropic
exponents from canonical simulations.  This idea is sketched
in Fig.~\ref{fig:sketch} and can be implemented in various ways.
One can consider, for instance, a polymer loop divided in two sides by
a wall (Fig.~\ref{fig:sketch}(a)); the two sides exchange monomers via
sufficiently small holes such that the knots cannot pass through. The
exchange can also occur through a fictitious ``wormhole''~\cite{zandi03},
as shown in Fig.~\ref{fig:sketch}(b). The polymers at the two sides
of the wall or those exchanging monomers through the wormhole do not
interact with each other. When exchanging monomers the length of each
loop fluctuates, while the total length is fixed to a constant $L$. The
method~\cite{zandi03} is based on the analysis of the equilibrium
distribution of lengths of the two sides. For ordinary polymers
one expects that the length $l$ of one polymer ring is distributed
according to
\begin{equation}
p(l) \sim Z^{(1)}_l Z^{(2)}_{L-l} \sim \mu^L l^{\theta_1} (L-l)^{\theta_2}
\label{pl}
\end{equation}
where the two $Z$'s are the loop partition functions given in
Eq.~(\ref{Zl_saw}). The main point is that the dependence on $\mu$
in Eq.~(\ref{pl}) is irrelevant as $L$ is fixed, whereas from the
analysis of the shape of the probability $p(l)$ as a function of $l$
it is possible to fit the values of the entropic exponents $\theta_1$ and
$\theta_2$ of the two loops~\cite{zandi03}.

\subsubsection{Entropic competition without a wall}

\begin{figure}[t]
\begin{center}
\includegraphics[angle=0,width=8cm]{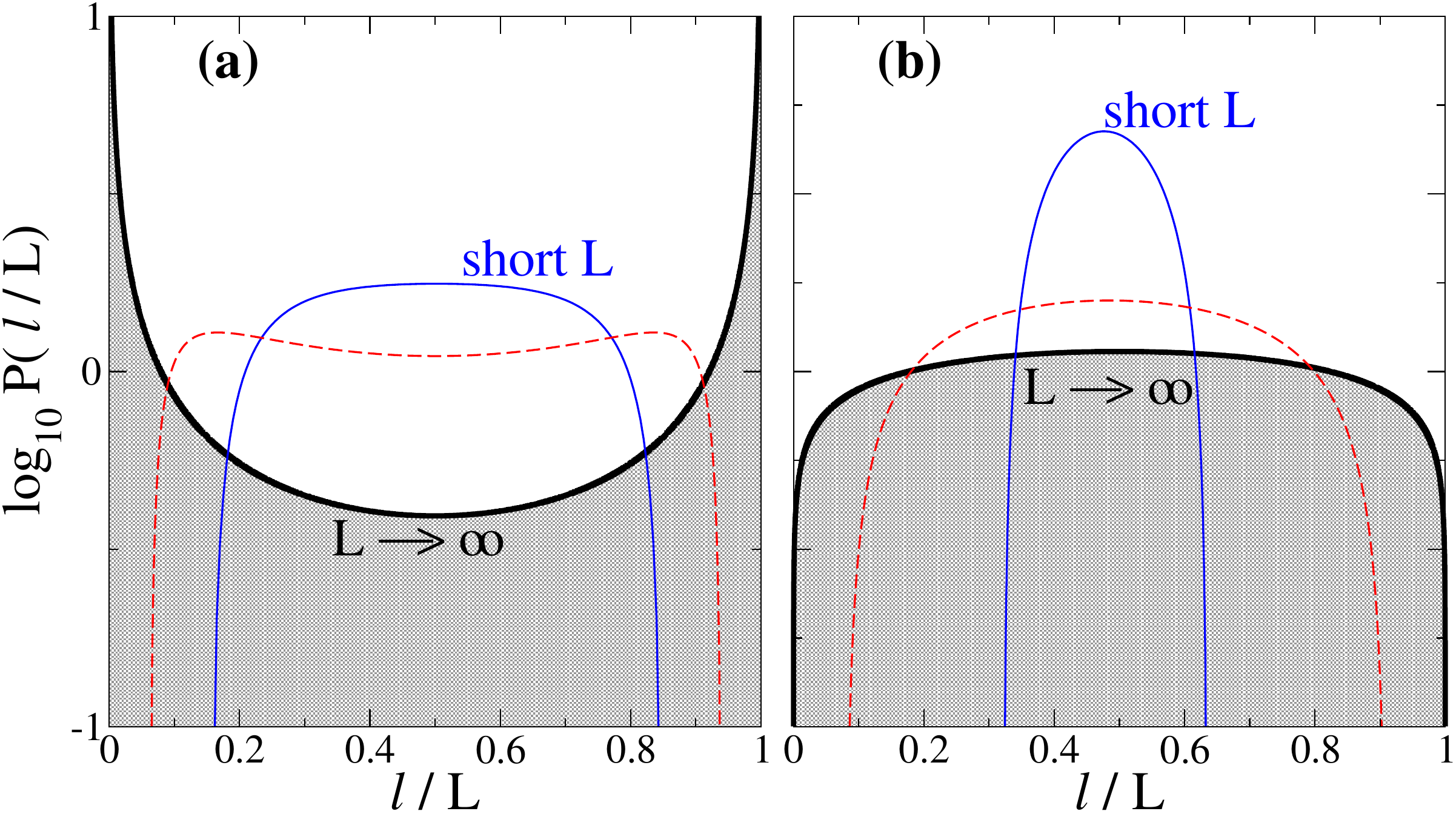}
\end{center}
\caption{(Color online) Examples of probability distributions of loop lengths for
two polymer loops exchanging monomers as in Fig.~\ref{fig:sketch}(b),
in the case of negative (a) and positive (b) entropic exponent $\theta$
for both loops.
In (a) the competition is between two $3_1$ knots, the
thick line (boundary of the shaded area) is the distribution 
for the limit of long $L$ while the other ones are for two short $L$'s.  
In (b) the competition is between a $3_1\#3_1$ knot and a $3_1\#4_1$ knot, 
with the same notation.}
\label{fig_new}
\end{figure}

We first consider the case depicted in Fig.~\ref{fig:sketch}(b), and we 
discuss a few representative examples. If the two loops both have negative 
entropic exponents ($\theta_1, \theta_2 < 0$), 
then one expects a $p(l)$ as depicted in
Fig.~\ref{fig_new}(a) (thick line and shaded area), whereas the case $\theta_1, \theta_2 > 0$ is
depicted in Fig.~\ref{fig_new}(b) (same notation;
in these figures, for convenience we show the distribution $p(l/L)$, which
is just a rescaling of $p(l)$).
The thin lines in Fig.~\ref{fig_new}(a) and (b) show sketches of finite-$L$
distributions of $p(l)$ for increasing $L$:
particularly interesting is the scenario depicted in 
Fig.~\ref{fig_new}(a), which shows a drastic change
of the shape of the distribution from a finite $L$ to the the limit
$L\to \infty$. We will discuss here how some of these features can be
understood from the analysis of the stored length densities $\rho^{(1)}_m$
and $\rho^{(2)}_m$ of the competing loops.

Let us consider $p(m)$, the probability of finding $m$ monomers in one
of the two entropically competing ELP loops. 
This quantity scales as
\begin{equation}
p(m)  \sim  \widetilde{Z}_m^{(1)} \widetilde{Z}_{N-m}^{(2)}
\label{maxpm}
\end{equation}
where the two $\widetilde{Z}$s are the partition functions of the two competing ELPs
at fixed monomer numbers $m$ and with fluctuating lengths. To find the most
probable value of the monomer number $m^*$ observed in the entropic
competition setup, we maximize the entropy
\begin{equation}
S_m = k_B \log \widetilde{Z}_m^{(1)} + k_B \log \widetilde{Z}_{N-m}^{(2)}
\label{Sm}
\end{equation}
($k_B$ is the Boltzmann constant).
The partition functions of ELPs in Eq.~(\ref{zm}) are expressed as a sum 
over all lengths $0 \leq m \leq l$. The sum is however dominated by 
a characteristic value of $l^*(m)$ obtained from the condition 
\begin{equation}
\frac{\partial \widetilde{Z}_{m,l}}{\partial l}\bigg|_{l=l^*(m)} = 0
\label{dZmldl=0}
\end{equation}
where we defined 
\begin{equation}
\widetilde{Z}_{m,l} \equiv \left( m \atop l \right) u^l Z_l.
\end{equation}
Now assuming that $\widetilde{Z}_m$ is dominated by a single value of
$l^*(m)$ we can compute the total derivative in $m$ of $\log \widetilde{Z}_m$ as
\begin{eqnarray}
\frac{d \log \widetilde{Z}_m}{d m} &\approx& 
\frac{d \log \widetilde{Z}_{m,l^*(m)} }{d m}  =
\frac{\partial \log \widetilde{Z}_{m,l^*(m)}}{\partial m} \nonumber\\ 
&=& \frac{d}{dm} \log \frac{m!}{(m-l^*(m))!} = -\log \rho_m.
\label{dlogZdm}
\end{eqnarray}
In this derivation we used Eq.~(\ref{dZmldl=0}), so in the total
derivative with respect to $m$ we can ignore the $m$-dependence coming
from $l^* (m)$.  Combining Eqs.~(\ref{Sm}) and (\ref{dlogZdm}) we
find that the extremum of the entropy $S_m$ of the competing rings
is given by the value of $m^*$ for which
\begin{equation}
\rho^{(1)}_{m^*} = \rho^{(2)}_{N-m^*}
\end{equation}
To find $m^*$ one can plot $\rho_{m}^{(1)}$ and
$\rho_{N-m}^{(2)}$ vs.~$m$ and $N-m$ in the same graph: each intersection
point between the two curves is an extremum of $S_m$. To decide whether
this is a local maximum or minimum one analyzes the second derivative 
\begin{equation}
\frac{d^2 S_m}{ dm^2}\bigg|_{m^*}  \simeq 
- \frac{1}{\rho^{(1)}_{m^*}}\left[\frac{d\rho^{(1)}_{m}}{dm} 
- \frac{d\rho^{(2)}_{N-m}}{d(N-m)} \right]_{m^*} \;.
\label{der2}
\end{equation}

\begin{figure}[!tb]
\begin{center}
\includegraphics[width=8cm,angle=0]{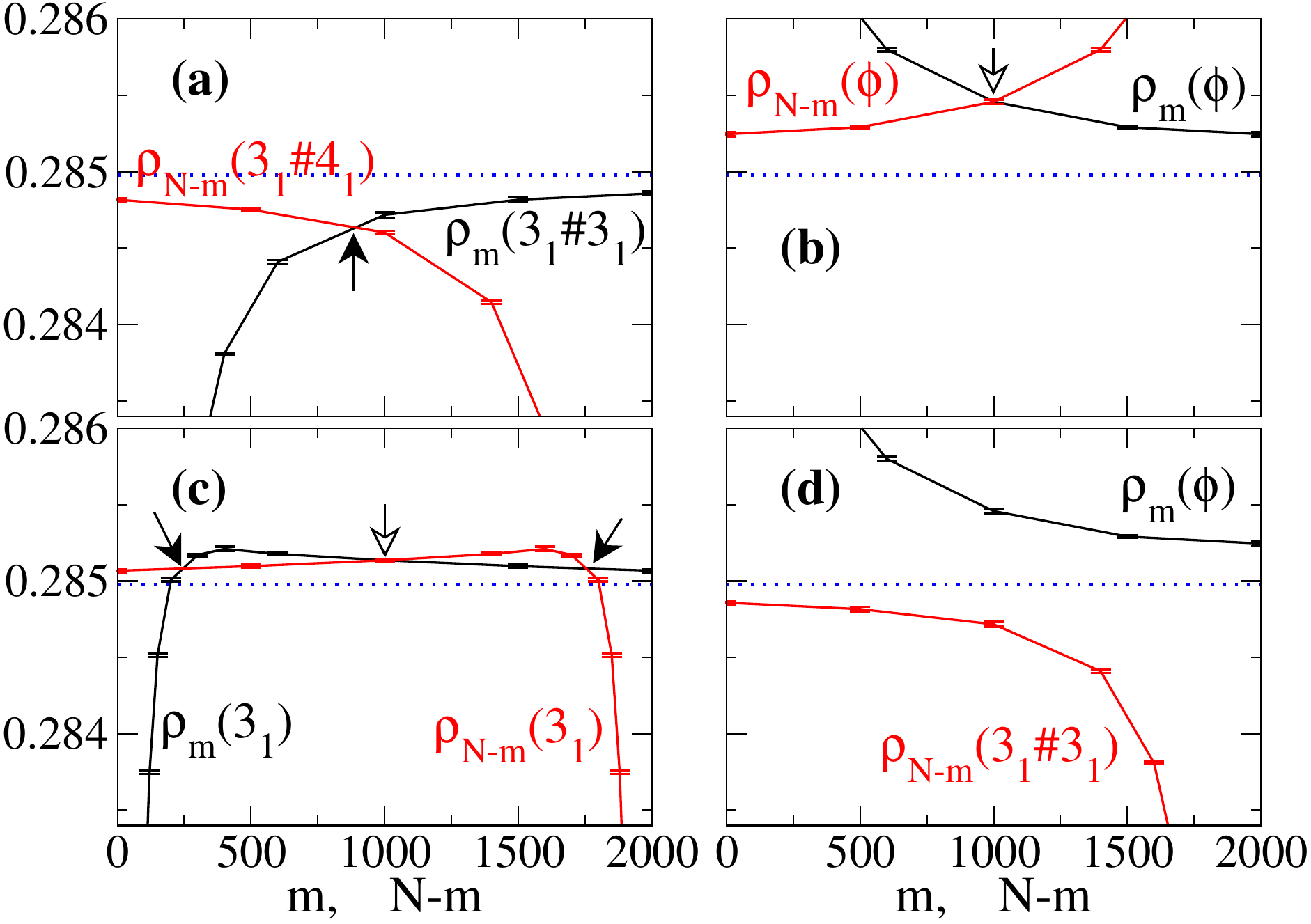}
\end{center}
\caption{(Color online) 
Plot of stored-length densities vs.~$m$ and $N-m$ in the entropic competition setup, 
for four different knotted chains: (a) $3_1\#3_1$ vs.~$3_1\#4_1$, (b) unknot vs.~unknot,
(c) $3_1$ vs.~$3_1$, and (d) unknot vs.~$3_1\#3_1$.
The intersection points of the densities are highlighted 
by arrows and correspond to local maxima (filled arrows) or minima (empty arrows) 
of the total entropy of the two competing loops.
Horizontal dotted lines indicate $\rho_\infty$.
\label{fig:comp}}
\end{figure}

In Fig.~\ref{fig:comp} we show some plots of the stored-length densities
for the two competing loops containing knots. The two loops have a total
number of monomers equal to $N=2000$ and the data are those shown in
Fig.~\ref{fig04}, but now plotted as function of $m$ and $N-m$. As seen
in the previous section, the stored length density can be non-monotonic in $m$ for
some knotted configurations, which can produce various scenarios where up
to three intersection points are possible.  

Figure~\ref{fig:comp}(a) shows the example of two competing double
knots. In this case there is a single intersection point and the
analysis of the first derivatives of $\rho$ shows that this point is
a local maximum for the entropy (Eq.~(\ref{der2})). The probability
distribution of monomers (or lengths in the canonical setup) will
have a single maximum at some intermediate $m^*$, as shown in the
example of Fig.~\ref{fig_new}(b). In the case of two unknotted loops
(Fig.~\ref{fig:comp}(b)) the intersection point $m^*$ is a minimum
for the entropy, hence the probability distribution for $m$ will be
maximal at the edges and minimal at $m^*$, as for the thick line in
Fig.~\ref{fig_new}(a).  The most interesting case is that of competition
between loops with non-monotonic $\rho$'s. This case is illustrated in
Fig.~\ref{fig:comp}(c). The three intersection points are a central local
minimum of the entropy enclosed by two local maxima.  The probability
distribution of lengths is like that depicted as a dashed line in 
Fig.~\ref{fig_new}(a). It
is easy to see that if the total number of monomers decreases (this
corresponds to shift one of the two $\rho$'s along the horizontal axis)
there will be only one intersection point.  This generates a probability
distribution with a single maximum for $m^*$ 
(thin dense line in Fig.~\ref{fig_new}(a)).  
Interestingly, the length
distributions obtained from Monte Carlo simulations~\cite{zandi03} of
competing off-lattice flexible rings with simple knots give, 
for sizes up to $200$ monomers, concave
distributions, contrary to the expectations of negative $\theta$'s from
the conjecture of Ref.~\cite{orla96}, which would instead correspond to
a convex (i.e. with a minimum in the middle) shape.  The non-monotonicity
in $m$ of the stored-length density $\rho_m$ explains this drastic change
in behavior in finite-size data.

To complete the discussion we consider next an example where no
intersection point is present (Fig.~\ref{fig:comp}(d)). In this case
one has to resort to the full form of $S_m$: from the scaling of
partition functions of SAPs, the probability of a
state with $m$ monomers on the side with no knots is expected to scale
as $m^{-3\nu}\sim m^{-1.76}$ with a cutoff at $m \lesssim N$.
It implies that the average length of the unknotted subchain
$\mean{m}_N\sim N^\alpha\sim N^{0.24}$ is weakly scaling with $N$, and
at least in this case the competition is clearly in favor of the side
with knots.  This reminds us that the full statistics given by $S_m$
would often be necessary to compute average quantities, and that the
maxima are only indicative elements.  Nevertheless, we have seen that the
density of stored length is a useful quantity for understanding the basic
properties of the entropy of competing knotted chains. In particular,
knowing it and its short-$N$ features helps to interpret the numerical
results and to distinguish preasymptotic scalings from asymptotic ones.

\subsubsection{Entropic competition with a wall}

Let us finally go back and reconsider briefly the entropic competition
of knots divided by a wall, as in Fig.~\ref{fig:sketch}(a). The main
difference is that the basic exponent of the unknotted chain should
be $\theta_s =  -d \nu+\sigma'_2$. The additional index
$\sigma'_2$ is connected to the constraint of having a monomer of a
loop confined close to a hard surface.  The formula is an application of Duplantier's
general theory of polymer networks~\cite{duplantier89:net,schafer92:net}.
We use this theory also to extract $\sigma'_2$ from the data in
Ref.~\cite{gaunt90}, obtaining $\sigma'_2\approx -0.95$.  This means that
$\theta_s\approx -2.71$. Again the full zoology of possible competitions
could be simply discussed by repeating the above reasoning, once data
of $\rho_m$ for ELPs close to a wall are generated. We reserve this
investigation for a future work.  Let us just note that the condition
$\theta_s+\pi_k>0$, associated with a single maximum of the entropy $S_m$
at $1\ll m^* \ll N$, is now met for a minimal number of prime knots
$\pi_k=3$ per loop, i.e.\ one more than we needed in the case without
the wall.  Thus, the wall separating the chains has somewhat the effect
of repelling entropically also the knots.

\section{Conclusions}

In this paper we studied the scaling properties of a class of polymers,
which we have referred to as elastic lattice polymers (ELPs). These
polymers can accumulate stored length along their backbone, by lifting
the self-avoidance condition for neighboring monomers.
The length $l$ of their backbone fluctuates, whereas
the total number of monomers $m$ is fixed. Differently from true grand
canonical polymers, however, the backbone length is bounded to $l \leq
m$, and fluctuates around  $\mean{l}$ with fluctuations $\sim \sqrt{m}$.

{
ELPs were used in the past to study the dynamics of polymer melts
\cite{heuk03} and pore translocation dynamics \cite{wolt06}, but their
equilibrium behavior has received little attention. In this work, we
used ELPs to investigate entropic exponents of knotted polymer rings.
The calculation of polymer entropic exponents from classical Monte Carlo
simulations of canonical self-avoiding rings is rather cumbersome:
one has either to employ complex grand canonical sampling, or to
resort to the so-called atmospheres method \cite{jans08} or entropic
competition methods \cite{zandi03}.  ELPs are well suited to this type
of problem. Firstly, the underlying Monte Carlo dynamics 
can be based solely on
local moves (thus conserving the knot topology) and the possibility
of accumulating length along the backbone facilitates the sampling of
different configurations compared to canonical self-avoiding rings.
The ELP explores new configurations through sliding moves along the
backbone. Secondly, we have shown that there is a natural variable
associated to ELPs which is the stored length }
density $\rho_m$ (see Eq.~\ref{def_rho}), which measures the average 
fraction of monomers accumulated on the backbone.  We have derived
an expansion for $\rho_m$ in the limit $m \to \infty$, where $\rho_m$
converges to a value that depends on the connectivity constant of
the ordinary lattice polymers. The next leading behavior is of the
order $\theta/m$, with $\theta$ the entropic exponent of the polymers.
This allows one to estimate entropic exponents from the scaling analysis
of $\rho_m$. As examples of application of this, we estimated entropic
exponents of swollen polymers in $d=2$ and $d=3$, and of polymers with
various types of knots.  Comparing the results with conjectured values
of these exponents, we find a clear agreement at least for the simplest
knot studied.  For more complex knots the agreement is only marginal,
due to finite-size effects quickly increasing with the knot complexity.

One of the advantages of the stored length analysis is that
correction-to-scaling effects are directly visible in $\rho_m$ vs. $1/m$
plots as they appear as deviations from a linear scaling behavior. Our
analysis showed that finite-size effects become stronger with the knot
complexity and with the number of knots. Similar result have been observed
by Janse van Rensburg and Rechnitzer~\cite{jans08}.  These authors
estimated the connectivity constant and entropic exponent of lattice
polymers via the atmosphere method~\cite{rechnitzer02:atmo},
where, roughly speaking, atmospheres are the loci where the polymer can
be expanded and contracted.  Interestingly, there is a similarity between
the scaling of the average atmospheres and that of the stored-length
density of ELPs discussed in this paper.

With simulations of ELPs we have shown how important are 
corrections to scaling in the statistics of knotted polymers:
their equilibrium properties in entropic competition 
can be understood from coexistence diagrams of stored
lengths of ELPs. The non-monotonicity of the stored length density as
function of $1/L$ explains some features of the competing rings observed
in canonical Monte Carlo simulations \cite{zandi03} 
which were poorly understood before.

Summarizing, the elastic lattice polymer is a simple model sharing 
critical exponents with the self-avoiding walk,  
but it has an additional ``elastic'' degree of freedom in its fluctuating length, 
which offers numerical advantages and additional theoretical tools to derive 
critical exponents of polymers. 
Thus, the ELP is a valid alternative to classical lattice models 
for studies in polymer physics.

%
%
%

\vspace{3mm}

\noindent {\bf Acknowledgments:}
G.B. and M.B. acknowledge hospitality from the Institute of Theoretical
Physics at the K.U.Leuven and financial support from the Research
Foundation-Flanders (FWO) Grant No. G.0311.08.  
M.B. acknowledges financial support from the University of Padua 
(Progetto di Ateneo n.~CPDA083702).


\end{document}